\documentclass[twocolumn,twoside,slac_two]{revtex4}
\usepackage{graphicx}
\usepackage{fancyhdr}
\begin{document}

\title{The July 2010 outburst of the NLS1 PMN J0948+0022}


\author{L. Foschini, G. Ghisellini, L. Maraschi, G. Tagliaferri, F. Tavecchio}
\affiliation{INAF Osservatorio Astronomico di Brera, 23807 Merate, Italy}
\author{Y.~Y. Kovalev, Yu.~A. Kovalev}
\affiliation{Astro Space Center of the Lebedev Physical Institute, 117997 Moscow, Russia}
\author{M.~L. Lister, J.~L. Richards}
\affiliation{Department of Physics, Purdue University, West Lafayette, IN 47907, USA}
\author{F. D'Ammando}
\affiliation{INAF Istituto di Radioastronomia, 40129 Bologna, Italy}
\author{D.~J. Thompson, D. Donato}
\affiliation{NASA Goddard Space Flight Center, Greenbelt, MD 20771, USA}
\author{A. Tramacere}
\affiliation{ISDC Data Centre for Astrophysics, CH-1290, Versoix, Switzerland}
\author{E. Angelakis, L. Fuhrmann, I. Nestoras}
\affiliation{Max-Planck-Institut f\"ur Radioastronomie, D-53121 Bonn, Germany}
\author{A. Falcone}
\affiliation{Department of Astronomy \& Astrophysics, Pennsylvania State University, PA 16802, USA}
\author{M. Hauser, S. Wagner}
\affiliation{Landessternwarte, Universit\"at Heidelberg, D 69117 Heidelberg, Germany}
\author{K. Mannheim, O. Tibolla}
\affiliation{University of W\"urzburg, 97074, W\"urzburg, Germany}
\author{W. Max-Moerbeck, V. Pavlidou, A.~C.~S. Readhead, M.~A. Stevenson}
\affiliation{Cahill Center for Astronomy and Astrophysics, California Institute of Technology, Pasadena, CA 91125, USA}
\author{A.~B. Pushkarev}
\affiliation{Crimean Astrophysical Observatory, 98409 Nauchny, Crimea, Ukraine}

\begin{abstract}
We report about the multiwavelength campaign on the Narrow-Line Seyfert 1 (NLS1)
Galaxy PMN~J0948$+$0022 ($z = 0.5846$) performed in 2010 July-September and triggered
by high activity as measured by \emph{Fermi}/LAT. The peak luminosity in the $0.1-100$~GeV energy band exceeded, for the first time in this type of source, the value of $10^{48}$~erg/s, a level comparable to the most powerful blazars. The comparison of
the spectral energy distribution of the NLS1 PMN~J0948$+$0022 with that of a
typical blazar -- like 3C 273 -- shows that the power emitted at gamma rays is
extreme.
\end{abstract}

\maketitle

\thispagestyle{fancy}


\section{INTRODUCTION}
Narrow-Line Seyfert 1 Galaxies (NLS1s) make a very peculiar class of active galactic nuclei (AGN). They were discovered in eighties (see \cite{POGGE} for a recent review) because of their difference in the optical spectra with respect to the classical Seyfert active nuclei. Basically, the full-width half maximum (FWHM) of their broad permitted emission lines (e.g. H$\beta$) has values smaller than those of Seyfert 1s, with a drop in the distribution of the values of FWHM(H$\beta$) above $\sim 2000$~km/s. These ``narrower'' broad-lines -- yes, it is an oxymoron! -- are not due to some obscuration, as indicated by the detection of the FeII bump in the optical spectra of NLS1s. The FeII bump is an indicator of the direct view of the broad-line region (BLR), because it is observed only in Seyfert 1s and not in Seyfert 2s (e.g. \cite{DONG}). Among the typical characteristics of NLS1s, the most interesting are the relatively low mass of the central spacetime singularity ($\sim 10^{6-8}M_{\odot}$) and the high rate of the accretion disc (up to the Eddington limit). To date, it is not yet clear if these -- as well as other peculiar observational characteristics -- are the symptoms of a central engine intrinsically different from the other Seyferts or NLS1s are just the low black hole mass tail of the Seyfert distribution\footnote{The state of the art of the researches in this field can be found in the proceedings of the workshop {\it Narrow-Line Seyfert 1 Galaxies and Their Place in the Universe}, Milano, Italy, 4-6 April 2011: http://pos.sissa.it/cgi-bin/reader/conf.cgi?confid=126)}.

As the other Seyferts, NLS1s are generally radio quiet, but $\sim 7$\% of them exhibit a very compact radio core with high brightness temperature and flat or even inverted spectrum, hints of the presence of a relativistic jet \cite{DOI,ZHOU,KOMOSSA}. These hints were strengthened by the detection in 1H~0323$+$342 ($z=0.061$) of a hard tail in the X-ray spectrum extending to hard X-rays and emerging during periods of high optical/UV fluxes \cite{FOSCHINI4}. An early attempt to detect high-energy $\gamma$ rays from radio-loud NLS1s was performed in 2004 with the {\it Whipple} Cerenkov telescope \cite{FALCONE}, but without success.

\begin{figure*}[!t]
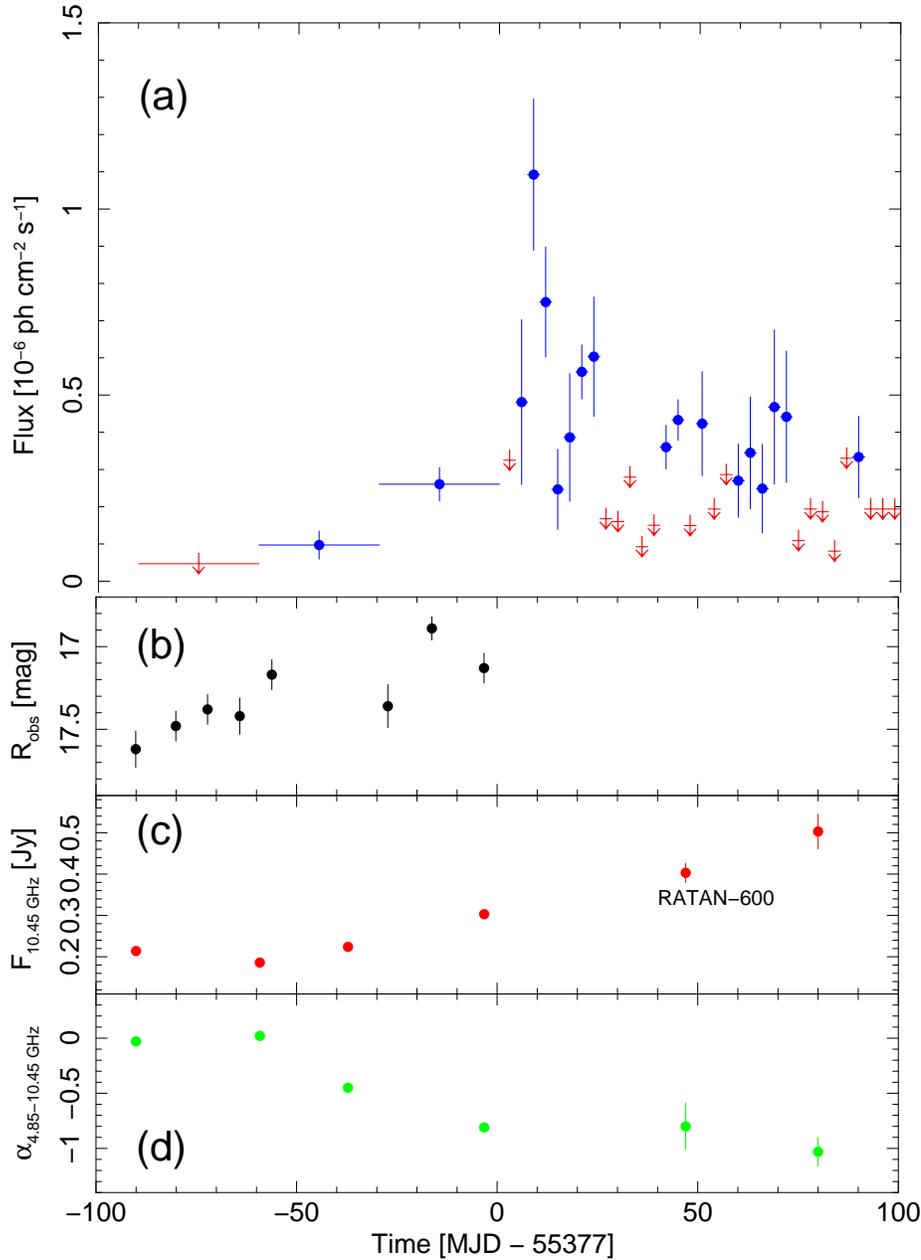

\centering
\includegraphics[angle=270,scale=0.5,clip, trim = 0 0 70 0]{curvagamma.ps}\\
\includegraphics[angle=270,scale=0.5]{ottico_specradio.ps}
\caption{{\it (a)} {\it Fermi}/LAT light curve of PMN~J0948$+$0022 in the $0.1-100$~GeV energy band with one-month time bin before of 2010 June 30 (MJD 55377) and 3-days time bin after that date. Upper limits (red arrows) are at $2\sigma$ level. {\it (b)} ATOM R filter observed magnitude. {\it (c)} Radio flux at $10.45$~GHz as observed from Effelsberg 100~m radiotelescope ({\it F-Gamma} program, see also Fig.~\ref{fig:effelsberg}). {\it (d)} Radio spectral index ($F_{\nu}\propto \nu^{-\alpha}$) between 5 and 10~GHz. The flux and spectral index at MJD 55424 (2010 August 16) is the average of the observations performed between 2010 August 13 and 29 with the RATAN-600 radio telescope at 11~GHz.} 
\label{fig:0948lc}
\end{figure*}

The definitive confirmation of the presence of powerful relativistic jets in NLS1s came with the Large Area Telescope (LAT) onboard the \emph{Fermi} satellite. After a few months of operations, LAT observed highly significant ($17\sigma$) emission of GeV photons from the NLS1 PMN~J0948$+$0022 ($z=0.5846$) \cite{DISCOVERY,FOSCHINI1}, which in turn was already known as strong radio-loud NLS1 \cite{ZHOU}. 

Soon after the early association of the $\gamma$-ray source with PMN~J0948$+$0022, made on a probabilistic basis, we started a multiwavelength campaign, which was performed between the end of 2009 March and the beginning of 2009 July \cite{MW2009}. The source displayed some activity at $\gamma$ rays with a peak of $\sim 4\times 10^{-7}$~ph~cm$^{-2}$~s$^{-1}$ ($0.1-100$~GeV) measured in one day on 2009 April 1. Then, the drop in the $\gamma$-ray emission was followed by a decrease of X-rays-to-optical flux and an increase of the radio flux after about less than two months. Particularly, {\it Swift}/UVOT recorded a significant change in the spectral slope of the optical emission during the decrease of the continuum. The observed coordinated broad-band variability confirmed that: 

\begin{enumerate}
\item[(a)] the $\gamma$-ray source discovered by {\it Fermi}/LAT is associated with the NLS1 PMN~J0948$+$0022;
\item[(b)] the continuum of the $\gamma$-NLS1 is dominated by the emission of a powerful relativistic jet viewed at small angles.
\end{enumerate}

Later, the observation of optical (V) polarization at $19$\% level with the KANATA telescope confirmed once more the above inferences \cite{KANATA}.

Other NLS1s have been detected in the GeV energy range by {\it Fermi} \cite{NLS1CLASS, FOSCHINI3}, thus indicating that a new class of $\gamma$-ray emitting AGN is emerging. One of the most important consequences of this discovery in our knowledge of relativistic jets is that it is now possible to study an unexplored range of black hole masses and accretion disc rates, which in turn is opening new horizons in our understanding of jets at all scales (see \cite{FOSCHINI3,FOSCHINI5,FOSCHINI6}). 

It is important to continue the monitoring of these sources in order to understand their nature and to extend the sample of $\gamma$-ray emitting candidates. Therefore, several monitoring programs were started and are currently ongoing. Again in 2010 July, PMN~J0948$+$0022 exploded in an intense outburst at $\gamma$ rays, reaching a peak luminosity of $\sim 10^{48}$~erg/s \cite{DONATO,FOSCHINI7}. The source was already under monitoring, but the high activity at $\gamma$ rays triggered more observations. Here we report about that campaign. Some early results were presented in \cite{FOSCHINI2} and a complete paper -- where more details are available -- has been recently published \cite{MW2010}. 

\begin{figure}[!t]
\centering
\includegraphics[angle=270,scale=0.32]{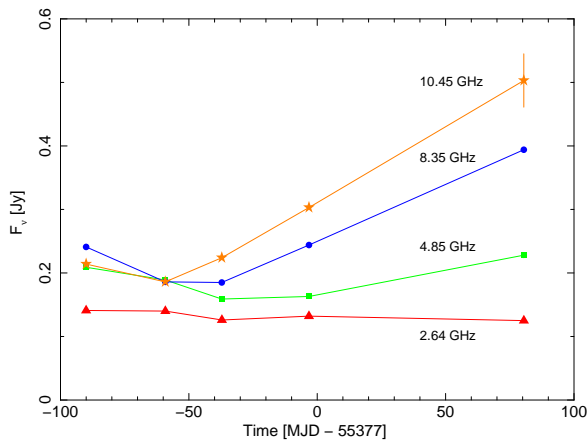}
\caption{Multifrequency light curves from Effelsberg 100~m radiotelescope ({\it F-Gamma} program). The different symbols indicate the measurements at different frequencies: red triangles, 2.64 GHz; green squares, 4.85~GHz; blue circles, 8.35~GHz; orange stars, 10.45~GHz. The reference time (MJD 55377) is 2010 June 30.} 
\label{fig:effelsberg}
\end{figure}

\section{FACILITIES INVOLVED}
After the $\gamma$-ray detection, PMN~J0948$+$0022 has been monitored by some facilities and, therefore, it was possible to reconstruct the multiwavelength behavior before and after the 2010 July outburst. There were some problems after the end of June, because the source position projected in the sky was apparently very close to the Sun, making it difficult the optical-to-X-ray observations. Anyway, we succeeded to collect sufficient data to do some important inferences on the nature of this source.

The data gathered to study the 2010 July outburst were from:

\begin{itemize}
\item {\it Fermi}/LAT, $0.1-100$~GeV energy band; the satellites continuously scans the $\gamma$ ray sky every two orbits (three hours). 

\item {\it Swift}, with its X-ray Telescope (XRT) operating in the $0.3-10$~keV energy band and its UltraViolet Optical Telescope (UVOT), equipped with the filters V, B, U, UVW1, UVM2, and UVW2. Only one short snapshot was possible and it was performed on 2010 July 3 (ObsID 00038394002), in the framework of the project {\it Swift/XRT Monitoring of Fermi/LAT sources of interest}\footnote{http://www.swift.psu.edu/monitoring/} at the Penn State University. The exposure on XRT was $\sim 1.6$~ks and the data from all the six UVOT filters were available.

\item Optical data in the R filter were taken with the Automatic Telescope for Optical Monitoring for HESS (ATOM), located in Namibia. 

\item Multifrequency radio observations ($2-43$~GHz, Fig.~\ref{fig:effelsberg}) were performed with the single-dish 100~m radiotelescope of Effelsberg ({\it F-Gamma} program\footnote{http://www.mpifr-bonn.mpg.de/div/vlbi/fgamma/fgamma.html}, \cite{FGAMMA,FGAMMA2}). PMN~J0948$+$0022 is continuously monitored on monthly basis since the discovery of GeV emission with {\it Fermi}.

\item Multifrequency radio observations ($5-22$~GHz) were performed in the period 2010 August 13-26 by using the RATAN-600 telescope \cite{KOVALEV}.

\item Radio observations at 15~GHz are continuously performed with the Owens Valley Radio Observatory (OVRO) within a program of monitoring of {\it Fermi} sources \cite{OVRO}.

\item VLBA high-resolution ($\sim$~mas scale) radio observations at 15~GHz within the {\it MOJAVE} Project\footnote{http://www.physics.purdue.edu/astro/MOJAVE/index.html} \cite{MOJAVE}. The source is monitored since the discovery of $\gamma$ rays.

\end{itemize}

\section{DISCUSSION}
Figure~\ref{fig:0948lc} displays four panels containing light curves at different frequencies, from radio to $\gamma$ rays. As in the 2009 campaign, the coordinated broad-band variability was dominated by the relativistic jet radiation.

It is possible to note that as the $\gamma$ ray emission became detectable (although with one month integration of data) about two months before the burst, there was an increase of the optical and radio flux, together with an inversion of the radio spectral index (see Fig.~\ref{fig:0948lc} and \ref{fig:effelsberg}). It is worth adding that VLBA osbervations indicated a $\sim 90^{\circ}$ swing in the electric vector position angle (EVPA) at some time between 2009 July 23 and 2009 December 10 (see Fig.~5 in \cite{MW2010}). A similar behavior has been already observed in high power blazars, such as PKS~1502$+$106 ($z=1.839$) \cite{1502}.  Perhaps, it is an indication of some arrangement of the jet structure to favor the radiative dissipation of the kinetic energy.

Another interesting feature observed during this outburst was the hardening of the $\gamma$ ray spectrum \cite{MW2010}. The photon index $\Gamma$ before and after the outburst was quite steep $\sim 2.7$, as expected from relativistic jets powered by external Compton (EC) processes. The EC needs of a nearby environment rich of photons, which in turn has the drawback to enhance the pair production probability and cutting the very high energy emission. At the peak of the emission, the $\gamma$-ray spectrum was harder, with $\Gamma \sim 2.5$. 

\begin{figure}[!t]
\centering
\includegraphics[angle=270,scale=0.32]{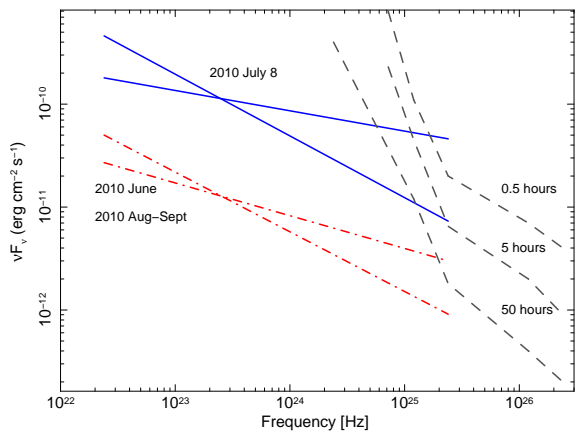}
\caption{$\gamma$-ray spectra during the outburst of PMN~J0948$+$0022 and perspectives of detection with CTA. The sensivity of CTA is plotted with grey dashed lines (with 0.5, 5, and 50 hours of exposure, from \cite{CTA}). The detection at the peak (2010 July 8) is indicated with blue continuous lines. The red dot-dashed lines refer to the spectrum measured in 2010 June and in the period August 1 and September 15, which are basically similar.} 
\label{fig:cta}
\end{figure}

A reanalysis of the LAT data with the most recent software ({\tt LAT Science Tools 9.23.1}), Instrument Response Function ({\tt P7SOURCE\_V6}), and background files ({\tt iso\_p7v6source.txt} and {\tt gal\_2yearp7v6\_v0.fits}), confirmed the hard spectrum, although not the change in the slope. In 2010 June (30 days of data), before the burst, the $0.1-100$~GeV flux and photon index $\Gamma$ were $(1.6\pm 0.3)\times 10^{-7}$~ph~cm$^{-2}$~s$^{-1}$ and $2.5\pm 0.1$, respectively. During the day of the burst (2010 July 8), the flux reached the value of $(1.3\pm 0.3)\times 10^{-6}$~ph~cm$^{-2}$~s$^{-1}$ with a photon index $2.4\pm 0.2$. The measurements done after the burst, by integrating data in the period August 1 - September 15 (45 days of data), resulted in the values of flux $(1.5\pm 0.2)\times 10^{-7}$~ph~cm$^{-2}$~s$^{-1}$ and photon index $2.4\pm 0.1$. What is important is the confirmation of the hardness of the spectrum, which can have important consequences on the possibility to detect these sources at even greater energies with ground-based Cherenkov telescopes (e.g. Cherenkov Telescope Array -- CTA -- \cite{CTA}, see Fig.~\ref{fig:cta}). Obviously, this depends on the possibility to have a spectral break at tens of GeV, but, if it is confirmed the similarity of behavior of $\gamma$-NLS1s with flat-spectrum radio quasars, it is reasonable to expect the occurrence of events like the case of PKS~B1222$+$216, detected at hundreds of GeV during a strong outburst \cite{FERMIPROC}. 

\begin{figure}[!t]
\centering
\includegraphics[angle=270,scale=0.5,clip, trim = 160 250 30 220]{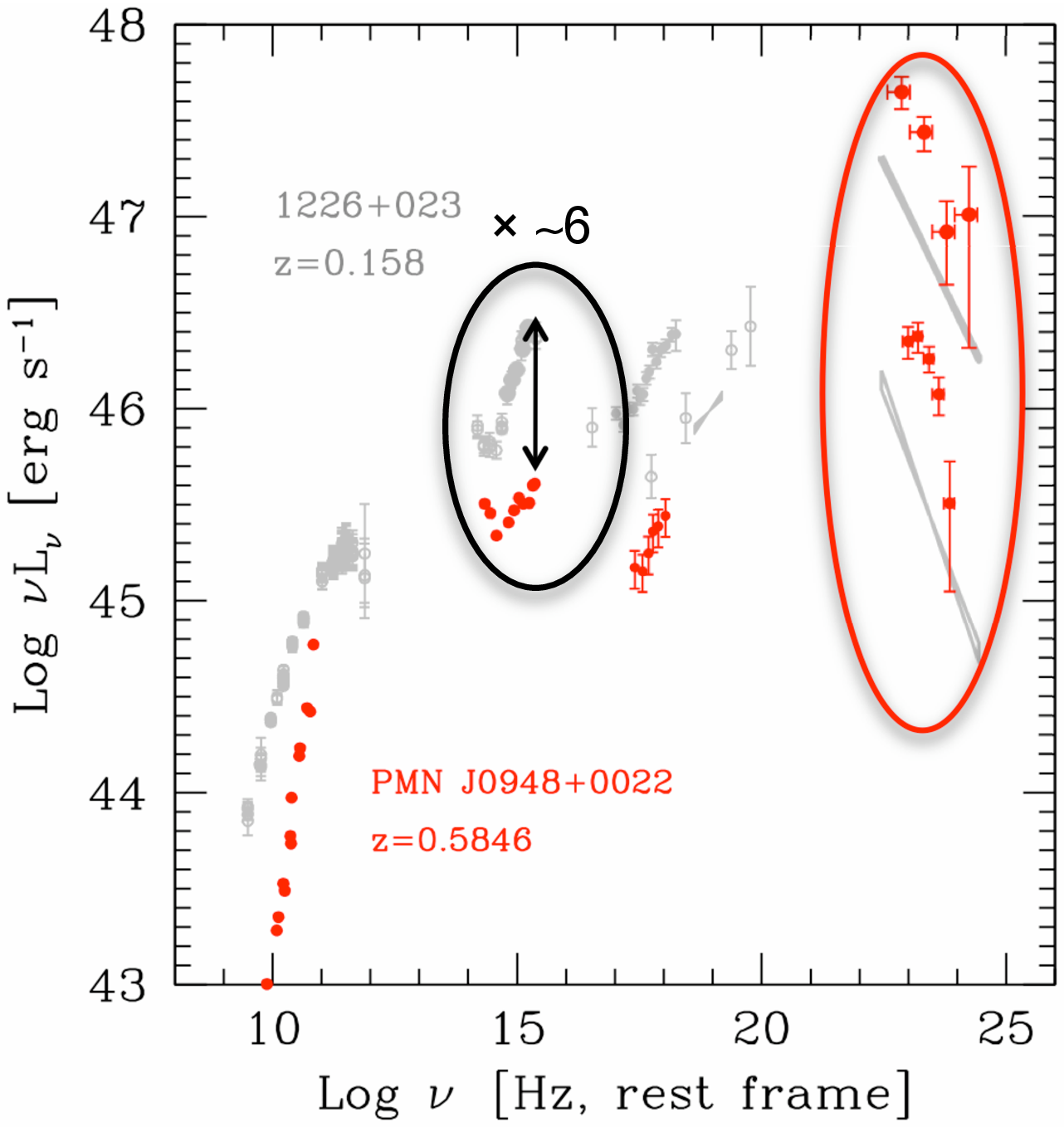}
\caption{Comparison of the SED of PMN~J0948$+$0022 (red points) and PKS~1226$+$023 (a.k.a. 3C~273) in grey. Adapted from \cite{MW2010}.} 
\label{fig:sed}
\end{figure}

More interesting inferences can be derived by comparing the spectral energy distribution (SED) of PMN~J0948$+$0022 with that of the archetypical blazar PKS~1226$+$023 (a.k.a. 3C~273, Fig.~\ref{fig:sed}). The radio-to-X-rays SEDs can be matched by multiplying the luminosities of PMN~J0948$+$0022 by a factor $\sim 6$, which in turn corresponds to the difference in mass of the two central black holes: $\sim 1.5\times 10^{8}M_{\odot}$ for the $\gamma$-NLS1 and $\sim 8\times 10^{8}M_{\odot}$ for the blazar. This rescaling, however, increases the separation at $\gamma$ rays, with the $\gamma$-NLS1 having the greatest power. This gap can be explained by a difference of the viewing angles, being that of PMN~J0948$+$0022 smaller than that of 3C~273. 

Later works supported these findings. The scaling in mass is justified by the fact that both sources have discs with high accretion rates ($\rightarrow$ radiation-pressure dominated) and hence the scaling of the Blandford-Znajek power is dependent only on the mass of the central black hole \cite{GOSH,MODERSKI}. The presence of two regimes (radiation- and gas-pressure dominated) is evident in the distribution of the jet power as a function of the accretion rate (see Fig.~3 of \cite{FOSCHINI5}) and both high-power blazars and $\gamma$-NLS1s occupy the region of sources with radiation-pressure dominated discs. 

\section{FINAL REMARKS}
The 2010 July outburst of the $\gamma$-ray NLS1 PMN~J0948$+$0022 showed once more the importance of multiwavelength campaigns in the understanding of sources with powerful relativistic jets. In this specific case, we noted:

\begin{enumerate}
\item[(a)] another similarity of the relativistic jet in PMN~J0948$+$0022 with those in blazars: the swing of the EVPA at radio frequencies before the outburst;

\item[(b)] the extreme efficiency of the jet power with respect to the mass of the central black hole ($\gamma$-NLS1s are small, but nasty). 

\item[(c)] the hardness of the $\gamma$-ray spectrum together with high flux during the outburst opens the possibility of a detection at hundreds of GeV with Cerenkov telescopes, which in turn will likely to be possible with future facilities, such as -- for example -- CTA.
\end{enumerate}

\bigskip
\begin{acknowledgments}
This research has made use of data obtained from the High Energy Astrophysics Science Archive Research Center (HEASARC), provided by NASA's Goddard Space Flight Center. 

This work has been partially supported by PRIN-MiUR 2007 and ASI Grant I/088/06/0. 

The OVRO 40~m monitoring program is supported in part by NASA (NNX08AW31G) and the NSF (AST-0808050). 

Based on observations with the 100-m telescope of the MPIfR (Max-Planck-Institut f\"ur Radioastronomie) at Effelsberg. Ioannis Nestoras is member of the International Max Planck Research School (IMPRS) for Astronomy and Astrophysics at the Universities of Bonn and Cologne. 

The \emph{Fermi} LAT Collaboration acknowledges support from a number of agencies and institutes for both development and the operation of the LAT as well as scientific data analysis. These include NASA and DOE in the United States, CEA/Irfu and IN2P3/CNRS in France, ASI and INFN in Italy, MEXT, KEK, and JAXA in Japan, and the K.~A.~Wallenberg Foundation, the Swedish Research Council and the National Space Board in Sweden. Additional support from INAF in Italy and CNES in France for science analysis during the operations phase is also gratefully acknowledged. 

This research has made use of data from the MOJAVE database that is maintained by the MOJAVE team (Lister et al. 2009). The MOJAVE project is supported under NSF grant AST-0807860 and NASA {\it Fermi} grant NNX08AV67G. 

RATAN-600 observations were supported in part by the Russian Foundation for Basic Research grant 08-02-00545. Y.~Y.~Kovalev was supported in part by the return fellowship of Alexander von Humboldt Foundation. 

\end{acknowledgments}


\bigskip 

\begin{thebibliography}{99}

\bibitem{CTA} CTA Consortium, ``Design Concepts for the Cherenkov Telescope Array CTA'', 2010, {\tt arXiv:1008.3703}

\bibitem{DOI} A. Doi et al., PASJ 58 (2006) 829.

\bibitem{DONG} X.~B. Dong et al., ApJ 721 (2010) L143.

\bibitem{FALCONE} A.~D. Falcone et al., ApJ 613 (2004) 710.

\bibitem{DISCOVERY} {\it Fermi}/LAT Coll. (A.~A. Abdo et al.), ApJ 699 (2009) 976.

\bibitem{MW2009} {\it Fermi}/LAT Coll. (A.~A. Abdo et al.), ApJ 707 (2009) 727.

\bibitem{NLS1CLASS} {\it Fermi}/LAT Coll. (A.~A. Abdo et al.), ApJ 707 (2009) L142.

\bibitem{DONATO} {\it Fermi}/LAT Coll. (D. Donato et al.), ATel 2733 (2010).

\bibitem{1502} {\it Fermi}/LAT Coll. (A.~A. Abdo et al.), ApJ 710 (2010) 810.

\bibitem{FOSCHINI1}	{\it Fermi}/LAT Coll. (L. Foschini et al.), ``Fermi/LAT Discovery of Gamma-Ray Emission from a Relativistic Jet in the Narrow-Line Seyfert 1 Quasar PMN~J0948$+$0022'', in: ``Accretion and ejection in AGN: a global view'', Como (Italy), 22-26 June 2009, ASP Conf. Series Vol. 427, p. 243 (2010).

\bibitem{FOSCHINI7} L. Foschini, ATel 2752 (2010).

\bibitem{FOSCHINI3} L. Foschini, ``Evidence of powerful relativistic jets in Narrow-Line Seyfert 1 Galaxies'', in: ``Narrow-Line Seyfert 1 Galaxies and Their Place in the Universe'', Milano (Italy), 4-6 April 2011, Proceedings of Science vol. NLS1, p. 024 (2011).

\bibitem{FOSCHINI5} L. Foschini, RAA {\bf 11} (2011) 1266.

\bibitem{FOSCHINI6} L. Foschini, {\tt arXiv:1107.2785}

\bibitem{FOSCHINI4} L. Foschini et al., Adv. Space Res 43 (2009) 889.

\bibitem{FOSCHINI2} L. Foschini et al., ``Relativistic Jets in Narrow-Line Seyfert 1'', in: ``Jets at all scales: Proceedings of the IAU Symposium 275'', Buenos Aires (Argentina), 13-17 September 2010, Proceedings IAU, Cambridge University Press, p. 176 (2011).

\bibitem{MW2010} L. Foschini et al., MNRAS 413 (2011) 1671.

\bibitem{FERMIPROC} L. Foschini et al., ``Short time scale variability at gamma rays in FSRQs and implications on the current models'', in: ``Proceedings of the Third Fermi Symposium'', Roma (Italy), 9-12 May 2011, eConf C110509, {\tt arXiv:1110.447}.

\bibitem{FGAMMA2} L. Fuhrmann et al., ``Simultaneous Radio to (Sub-) mm-Monitoring of Variability and Spectral Shape Evolution of potential GLAST Blazars'', in: ``The First GLAST Symposium'', Stanford (CA, USA), 5-8 February 2007, AIP Conf. Proc. vol. 921, p. 249 (2007).

\bibitem{FGAMMA} L. Fuhrmann et al., ``Gamma-ray NLSy1s and 'classical' blazars: are they different at radio cm/mm bands?'', in: ``Narrow-Line Seyfert 1 Galaxies and Their Place in the Universe'', Milano (Italy), 4-6 April 2011, Proceedings of Science vol. NLS1, p. 026 (2011).

\bibitem{GOSH} P. Gosh \& M.~A. Abramowicz, MNRAS 292 (1997) 887.

\bibitem{KOMOSSA} S. Komossa et al., AJ 132 (2006) 531.

\bibitem{KOVALEV} Y.~Y. Kovalev et al., A\&AS 139 (1999) 545.

\bibitem{KANATA} Y. Ikejiri et al., PASJ 63 (2011) 639.

\bibitem{MOJAVE} M.~L. Lister et al., AJ 137 (2009) 3718.

\bibitem{MODERSKI} R. Moderski \& M. Sikora, MNRAS 283 (1996) 854.

\bibitem{POGGE} R. Pogge, ``A quarter century of Narrow-Line Sefert 1s'', in: ``Narrow-Line Seyfert 1 Galaxies and Their Place in the Universe'', Milano (Italy), 4-6 April 2011, Proceedings of Science vol. NLS1, p. 002 (2011).

\bibitem{OVRO} J. Richards et al., ``15 GHz Monitoring of Gamma-ray Blazars with the OVRO 40 Meter Telescope in Support of Fermi'', in: ``The Second Fermi Symposium'', Washington DC (USA), 2-5 November 2009, eConf C0911022, {\tt arXiv:0912.3780}.

\bibitem{ZHOU} H.~Y. Zhou et al., ApJ 584 (2003) 147.

\end{thebibliography}
\end{document}